\documentclass[%
reprint,
showpacs,
showkeys,
nofootinbib,
nobibnotes,
 amsmath,
 amssymb,
aps, 
superscriptaddress,
]{revtex4-1}
\usepackage{graphicx}
\usepackage{epstopdf}
\usepackage{amssymb}
\usepackage{color}
\newcommand{\keVnr}{\ensuremath{\textrm{keV}_\textrm{nr}}} 
\newcommand{\keVee}{\ensuremath{\textrm{keV}_{\textrm{ee}}}} 

\newcommand{\scix}[3][]  
{ 
 \ifthenelse{\equal{#2}{}} 
  {} 
  { 
   {#2} 
  } 
 \ifthenelse{\equal{#2}{}\OR\equal{#3}{}} 
  {} 
  { 
   {\hspace{-0em} \times \hspace{-0em}} 
  } 
 \ifthenelse{\equal{#3}{}} 
  {} 
  {10^{#3}}   
 \ifthenelse{\equal{#1}{}}
 {}
 {\,\mathrm{#1}} 
}

\usepackage{dcolumn}
\usepackage{bm}
\usepackage{color}

\begin{document}

\affiliation{Brown University, Dept. of Physics, 182 Hope St., Providence RI 02912, USA}
\affiliation{Case Western Reserve University, Dept. of Physics, 10900 Euclid Ave, Cleveland OH 44106, USA}
\affiliation{Harvard University, Dept. of Physics, 17 Oxford St., Cambridge MA 02138, USA}
\affiliation{Imperial College London, High Energy Physics, Blackett Laboratory, London SW7 2BZ, UK}
\affiliation{Lawrence Berkeley National Laboratory, 1 Cyclotron Rd., Berkeley CA 94720, USA}
\affiliation{Lawrence Livermore National Laboratory, 7000 East Ave., Livermore CA 94550, USA}
\affiliation{LIP-Coimbra, Department of Physics, University of Coimbra, Rua Larga, 3004-516 Coimbra, Portugal}
\affiliation{South Dakota School of Mines and Technology, 501 East St Joseph St., Rapid City SD 57701, USA}
\affiliation{South Dakota Science and Technology Authority, Sanford Underground Research Facility, Lead, SD 57754, USA}
\affiliation{Texas A \& M University, Dept. of Physics, College Station TX 77843, USA}
\affiliation{University College London, Department of Physics and Astronomy, Gower Street, London WC1E 6BT, UK}
\affiliation{University of California Berkeley, Department of Physics, Berkeley CA 94720, USA}
\affiliation{University of California Davis, Dept. of Physics, One Shields Ave., Davis CA 95616, USA}
\affiliation{University of California Santa Barbara, Dept. of Physics, Santa Barbara, CA, USA}
\affiliation{University of Edinburgh, SUPA, School of Physics and Astronomy, Edinburgh, EH9 3JZ, UK}
\affiliation{University of Maryland, Dept. of Physics, College Park MD 20742, USA}
\affiliation{University of Rochester, Dept. of Physics and Astronomy, Rochester NY 14627, USA}
\affiliation{University of South Dakota, Dept. of Physics, 414E Clark St., Vermillion SD 57069, USA}
\affiliation{Yale University, Dept. of Physics, 217 Prospect St., New Haven CT 06511, USA}

\title{First results from the LUX dark matter experiment at the Sanford Underground Research Facility}

\author{D.S.~Akerib} 
\affiliation{Case Western Reserve University, Dept. of Physics, 10900 Euclid Ave, Cleveland OH 44106, USA}

\author{H.M.~Ara\'{u}jo} 
\affiliation{Imperial College London, High Energy Physics, Blackett Laboratory, London SW7 2BZ, UK}

\author{X.~Bai} 
\affiliation{South Dakota School of Mines and Technology, 501 East St Joseph St., Rapid City SD 57701, USA}

\author{A.J.~Bailey} 
\affiliation{Imperial College London, High Energy Physics, Blackett Laboratory, London SW7 2BZ, UK}

\author{J.~Balajthy} 
\affiliation{University of Maryland, Dept. of Physics, College Park MD 20742, USA}

\author{S.~Bedikian} 
\affiliation{Yale University, Dept. of Physics, 217 Prospect St., New Haven CT 06511, USA}

\author{E.~Bernard} 
\affiliation{Yale University, Dept. of Physics, 217 Prospect St., New Haven CT 06511, USA}

\author{A.~Bernstein} 
\affiliation{Lawrence Livermore National Laboratory, 7000 East Ave., Livermore CA 94550, USA}

\author{A.~Bolozdynya} 
\affiliation{Case Western Reserve University, Dept. of Physics, 10900 Euclid Ave, Cleveland OH 44106, USA}

\author{A.~Bradley} 
\affiliation{Case Western Reserve University, Dept. of Physics, 10900 Euclid Ave, Cleveland OH 44106, USA}

% ...


\begin{abstract}
The Large Underground Xenon (LUX) experiment is a dual-phase xenon time-projection chamber operating at the Sanford Underground Research Facility (Lead, South Dakota).  The LUX cryostat was filled for the first time in the underground laboratory in February~2013. We report results of the first WIMP search dataset, taken during the period April to August~2013, presenting the analysis of 85.3~live-days of data with a fiducial volume of 118~kg.  A profile-likelihood analysis technique shows our data to be consistent with the background-only hypothesis, allowing 90\% confidence limits to be set on spin-independent WIMP-nucleon elastic scattering with a minimum upper limit on the cross section of $7.6 \times 10^{-46}$~cm$^{2}$ at a WIMP mass of 33~GeV/c$^2$.  We find that the LUX data are in disagreement with low-mass WIMP signal interpretations of the results from several recent direct detection experiments.

\end{abstract}

          
\maketitle

Convincing evidence for the existence of particle dark matter is derived from observations of the universe on scales ranging from the galactic to the cosmological~\cite{Blumenthal:1984bp,Davis:1985,Clowe:2006}. Increasingly detailed studies of the Cosmic Microwave Background anisotropies have implied the abundance of dark matter with remarkable precision~\cite{Hinshaw:2012aka,Ade:2013zuv}. One favored class of dark matter candidates, the Weakly Interacting Massive Particle (WIMP), may be amenable to direct detection in laboratory experiments through its interactions with ordinary matter~\cite{PhysRevD.31.3059, Feng:2010}. The WIMPs that constitute our galactic halo would scatter elastically with nuclei, generating recoil energies of several keV.

We report here the first results from the Large Underground Xenon (LUX) experiment, currently operating 4850~feet below ground (4300~m~w.e.)~at the Sanford Underground Research Facility (SURF)~\cite{lab1,lab2} in Lead, South Dakota.  Fluxes of cosmic-ray muons, neutrons and $\gamma$-rays at SURF have been published elsewhere~\cite{lab3}.  Inside the cavern, a 7.6~m diameter by 6.1~m tall cylindrical water tank provides shielding to the detector.  These large reductions in external radiation provide the low-background environment required for the rare event search.  

The LUX detector holds 370~kg of liquid xenon, with 250~kg actively monitored in a dual-phase (liquid-gas) time-projection chamber (TPC) measuring 47~cm in diameter and 48~cm in height (cathode-to-gate)~\cite{LUXNimPaper}. Interactions in the liquid produce prompt scintillation (S1) and ionization electrons that drift in an applied electric field~\cite{araujoReview}.  Electrons are extracted into the gas, where they produce electroluminescence (S2). S1 and S2 signals are used to reconstruct the deposited energy and their ratio is used to discriminate nuclear recoils (NR) from electron recoils (ER).  Light signals are detected via two arrays of 61~photomultiplier tubes (PMTs), one array above the active region in the gas and one below it in the liquid~\cite{LUXPMTPaper}.  During this search, three PMTs were left unbiased, two in the top array and one in the bottom (one PMT was grounded and the others produced an abnormal increase in event rate).  The ($x$,$y$) position of an interaction is determined from localization of the S2 signal in the top PMT array, with the difference in time between the S1 and S2 representing event depth.  The ($x$,$y$) position resolution for small S2 signals (such as those in the WIMP search region in terms of both energy and fiducial volume) is 4--6~mm, and even better at higher energies. S2 pulse areas measured from the bottom PMT array alone (S2$_b$) are used in later analysis, avoiding events leaking into the signal region due to uncollected S2 light from the deactivated PMTs in the top array.

Throughout the WIMP search, the xenon vessel was thermally isolated with an outer vacuum vessel providing thermal stability of $\Delta T{<}0.2$~K, pressure stability $\Delta P/P{<}1\%$ and liquid level variation of ${<}0.2$~mm~\cite{PatrickPhelpsThesis} (measured from stability of S2 width).  An electric field of 181~V/cm was applied across the WIMP target region providing a measured average electron drift velocity of $1.51 \pm 0.01$~mm/$\mu$s.  Above the drift region, a field of 6.0~kV/cm is applied to the gas (3.1~kV/cm in the liquid), producing a best-fit electron extraction efficiency of $0.65 \pm 0.01$.   The distribution of the number of S2 photoelectrons observed for each extracted electron has a mean of 24.6 and an rms variation of 7.0. If only (S2$_b$) is considered, the mean is 10.4 and the rms is 4.5. 

Purification of the xenon, circulating through a hot-zirconium getter (229~kg/day), resulted in mean electron drift-lengths, before capture by electronegative impurities, between $87{\pm}9$ and $134{\pm}15$~cm during WIMP search.  

The data acquisition (DAQ) threshold is set such that ${\gtrsim}95\%$ of all single photoelectron (phe) pulses in each PMT are recorded to disk~\cite{LUXdaq,JeremyChapmanThesis}.  A digital trigger identifies events for further analysis, with non-adjacent PMTs grouped together into 16~trigger channels. The trigger requires that at least 2 of these channels have greater than 8~phe within a 2~$\mu$s window, with a trigger efficiency ${>}99\%$ for S2 signals above the analysis threshold of 200~phe. Every pulse of light digitized by the DAQ within $\pm 500$~$\mu$s of the trigger time (324~$\mu$s~maximum drift time) was allocated to a triggered event for further analysis, ensuring that corresponding S1 and S2 pulses can always be associated. Additionally, data between triggered events are retained to verify that the detector is quiet in the period leading up to, and following, the events. 

This initial dark matter search consists of 85.3~live-days of WIMP search data acquired between April~21 and August~8,~2013. The live-time calculation accounts for the DAQ dead-time ($0.2\%$), a 1--4~ms trigger hold-off to prevent additional triggers following large S2 pulses ($2.2\%$), and exclusions for periods of detector instability ($0.8\%$).

A non-blind analysis was conducted on the 85.3~live-days of WIMP search data, where only a minimal set of data quality cuts, with high acceptance, was employed to reduce the scope for bias.  The low total event rate in the center of the detector minimizes the rate of misidentified ER background events.  For this initial analysis of the first low-background operation of the instrument, both the calibration and WIMP search data were used to understand and develop analysis algorithms.

Waveforms from each PMT are summed across all channels and then searched with pulse finding algorithms to select viable signals.  The identification of an S1 signal requires at least two PMTs to detect more than 0.25~phe each within 100~ns of each other.  The average dark count rate for each PMT in the array is 12~Hz.  An estimate of the rate of events where an accidental 2~phe dark count coincidence fakes an S1 preceding a valid S2-only event in the NR signal region is 1.2~nHz  (0.009~events in the search dataset)~\cite{JeremyChapmanThesis}. 

Events containing exactly one S1 within the maximum drift time (324~$\mu$s) preceding a single S2, representative of expected elastic scattering of WIMPs, are selected for further analysis.  Additionally, we require a raw S2 pulse size greater than 200~phe ($\sim$8~extracted electrons).  This excludes a small number of single-extracted-electron-type events (having poor event reconstruction) and those from the detector walls with small S2 signals (having poorly reconstructed positions).  The 200~phe threshold for S2 light was optimized by studying the efficiency of the reconstruction algorithms with the calibration data and by observing the background outside the WIMP search energy range. 

\begin{figure}[h]
\begin{center}
\includegraphics[width=0.48\textwidth,clip]{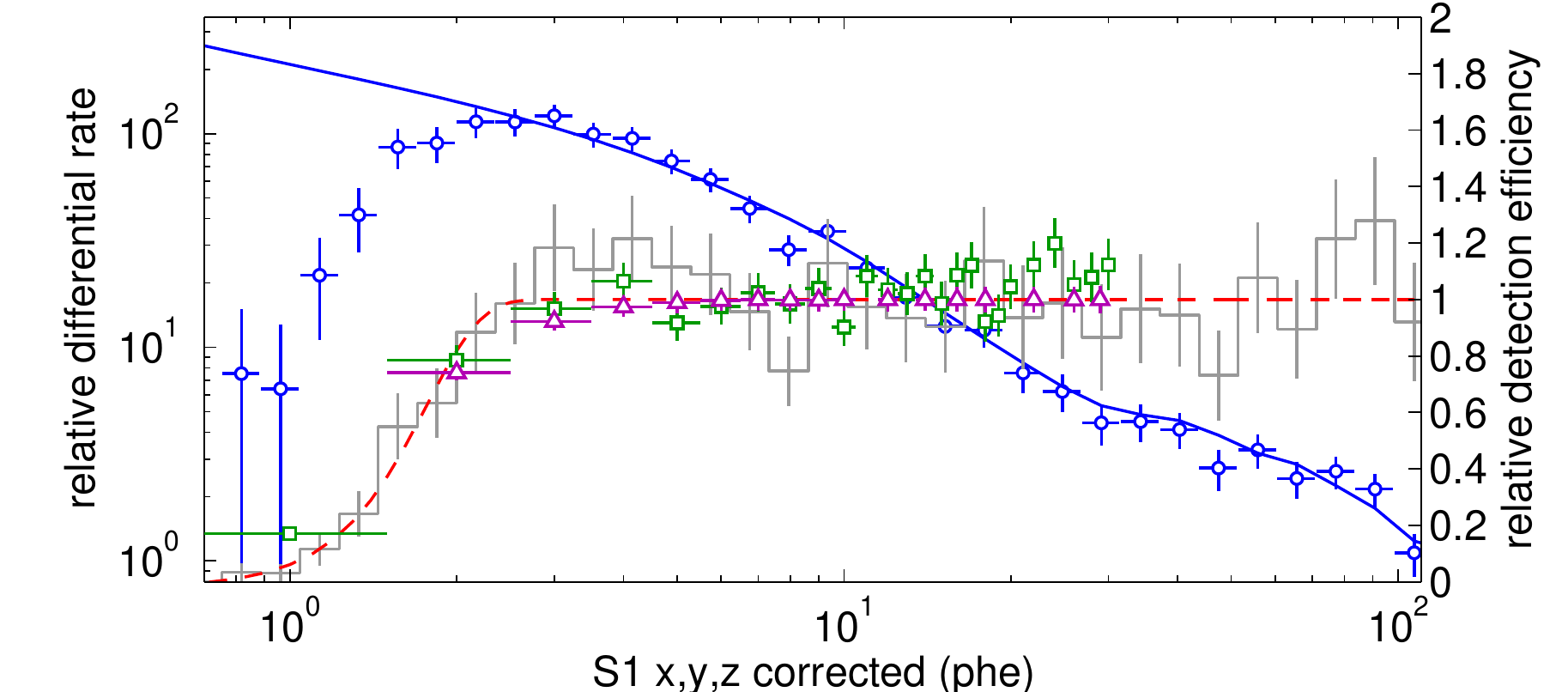}
\includegraphics[width=0.48\textwidth,clip]{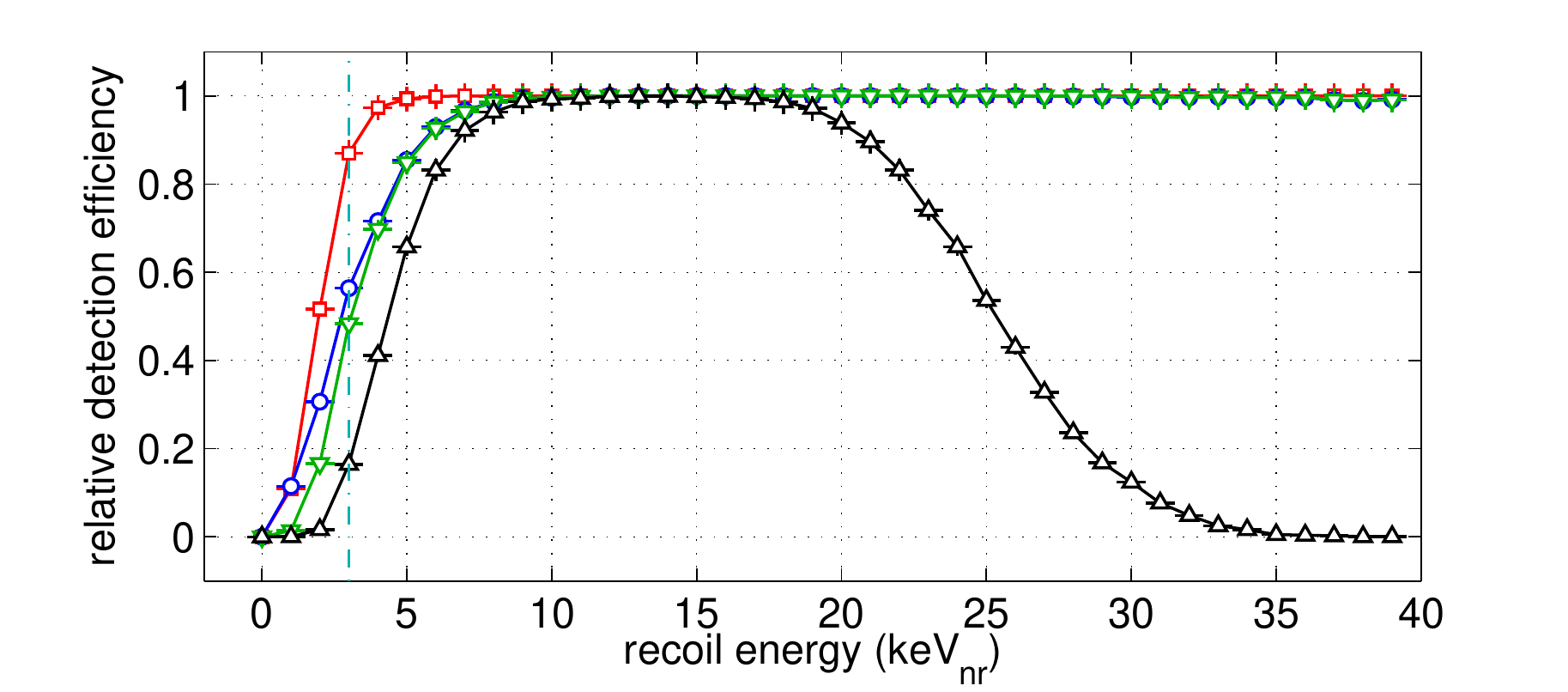}
\caption{\label{Efficiencies} {\it Top}:  Comparison of AmBe data (blue circles) with NEST simulations (blue line), showing excellent agreement above the 2~phe threshold (left axis).  The gray histogram and fitted dashed red line show the relative efficiency for detection of nuclear recoils from AmBe data (right axis).  Overlaid are the ER detection efficiency from tritium data (green squares), applied to the ER background model in the profile likelihood analysis, and the efficiency from full detector NR simulations treated as real data in terms of the digitized MC-truth S1 phe (purple triangles), applied to the WIMP signal model.  The efficiency calculation here does not include S1 or S2 area thresholds.  {\it Bottom}:  NR detection efficiency as a function of nuclear recoil energy for events with a corrected S1 between 2 and 30~phe and a S2 signal greater than 200~phe (black $\bigtriangleup$), the efficiency used directly in the profile likelihood analysis.  The efficiency for individually detecting an S2 (red $\Box$) or S1 (blue $\bigcirc$) signal (without the application of any analysis thresholds) are also shown, along with that after the single scatter requirement (green $\bigtriangledown$).  The cyan dashed line indicates the threshold in \keVnr{} below which we assume no light or charge response in the PLR calculation.}
\end{center}
\end{figure}

Single scatter ER and NR acceptance was measured with dedicated tritium ($\beta^-$), AmBe, and $^{252}$Cf (neutron) datasets.  Simulated NR event waveforms, generated with LUXSim~\cite{LUXSim,NESTmock}, were analyzed with the complete data processing framework to validate the analysis efficiencies measured with data. S2 finding efficiency is ${>} 99$\% above the analysis threshold of 200~phe.  The relative efficiency for NR detection is dominated by the S1 identification (shown in Fig.~\ref{Efficiencies}).  Absolute efficiency is estimated through visual inspection of waveforms from NR calibration data to be 98\%, which is in agreement with the value measured by an injection of tritiated methane of known activity.  All cuts and efficiencies combine to give an overall WIMP detection efficiency of 50\%~at 4.3~\keVnr{} (17\% at 3~\keVnr{} and $> 95\%$ above 7.5~\keVnr), shown in Fig.~\ref{Efficiencies}.  

\begin{figure}[h]
\begin{center}
\includegraphics[width=0.49\textwidth,clip]{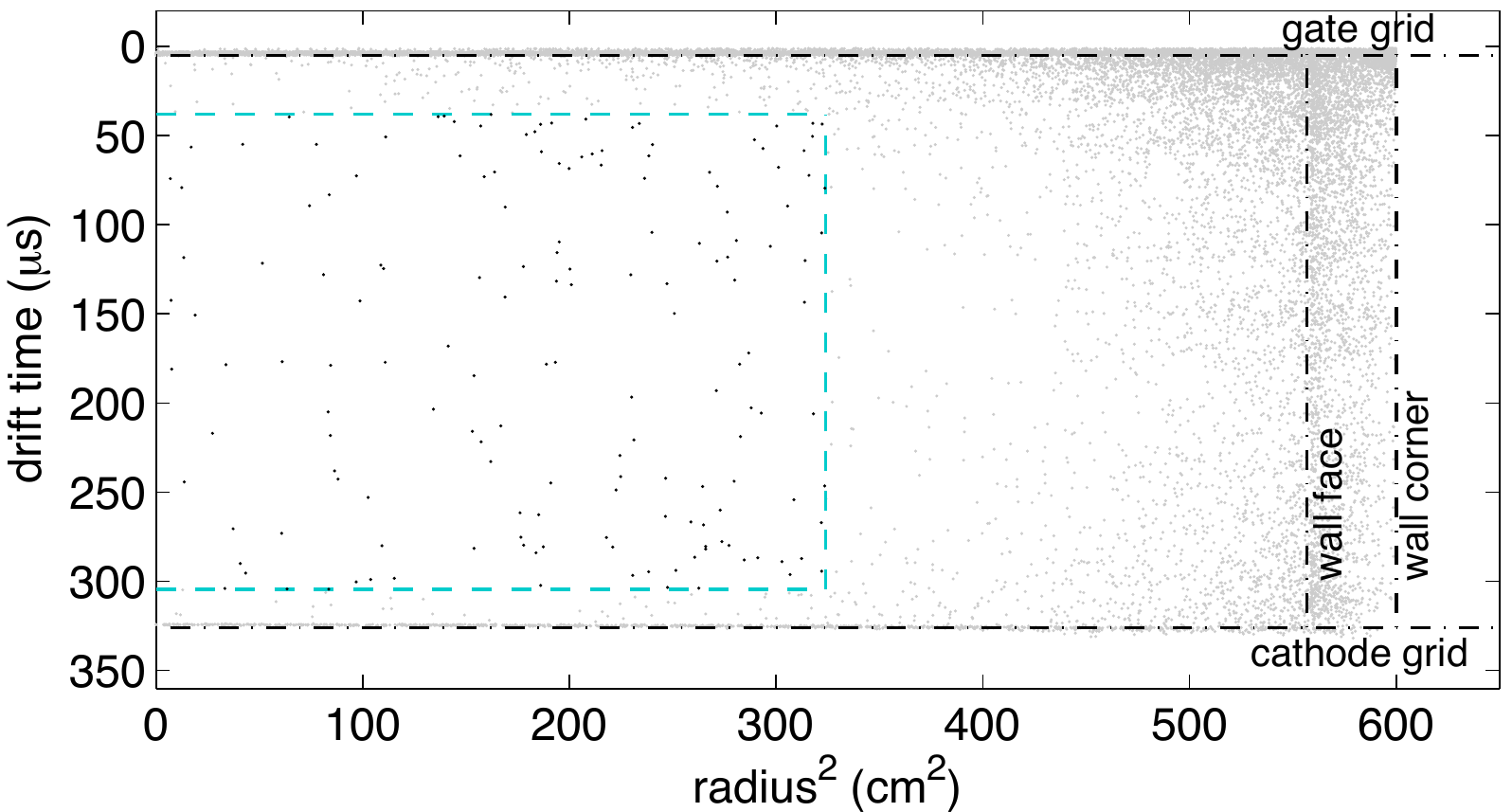}
\caption{\label{WSrz} Spatial distribution of all events with position-corrected S1 in the range 2-30~phe from the 85.3~live-days of WIMP search data.  The cyan dashed line indicates the fiducial volume.  The physical locations of the cathode and gate grids and the detector walls (where the vertical PTFE walls of the TPC form a dodecagon) are also shown.}  
\end{center}
\end{figure}

A radial fiducial cut was placed at 18~cm (Fig.~\ref{WSrz}), defined by the positions of decay products from Rn daughters implanted on the detector walls.  This population, primarily sub-NR band but intersecting the signal region at the lowest energies, is visible (along with other expected backgrounds) on the detector walls in Fig.~\ref{WSrz}. This cut was chosen by selecting those sub-NR band events outside of the WIMP search energy range (S1$>$30~phe).  In height, the fiducial volume was defined by a drift time between 38 and 305~$\mu$s to reduce backgrounds from the PMT arrays and electrodes. This cut was chosen by examining the event rate as a function of depth outside of the WIMP search energy range (S1$>$30~phe) and confirmed with Monte Carlo simulations. The fiducial target mass is calculated to be $118.3 \pm 6.5$~kg from assessment of the homogeneous tritium data, and confirmed from assessment of the homogeneous $^{83m}$Kr data, whose mono-energetic peak provides excellent tagging to monitor dispersal of the $^{83m}$Kr throughout the detector volume.

Periods of live-time with high rates of single electron backgrounds ($\gtrsim 4$~extracted electrons per 1~ms event window) are removed~\cite{ZIIse,PeterXe10S2only,ZIIIse}.  The associated loss of live-time is 0.8\% (measured from assessment of the full dataset, including non-triggered regions), primarily removing periods following large S2 pulses. 

\begin{figure}[h]
\begin{center}
\includegraphics[width=0.48\textwidth,clip]{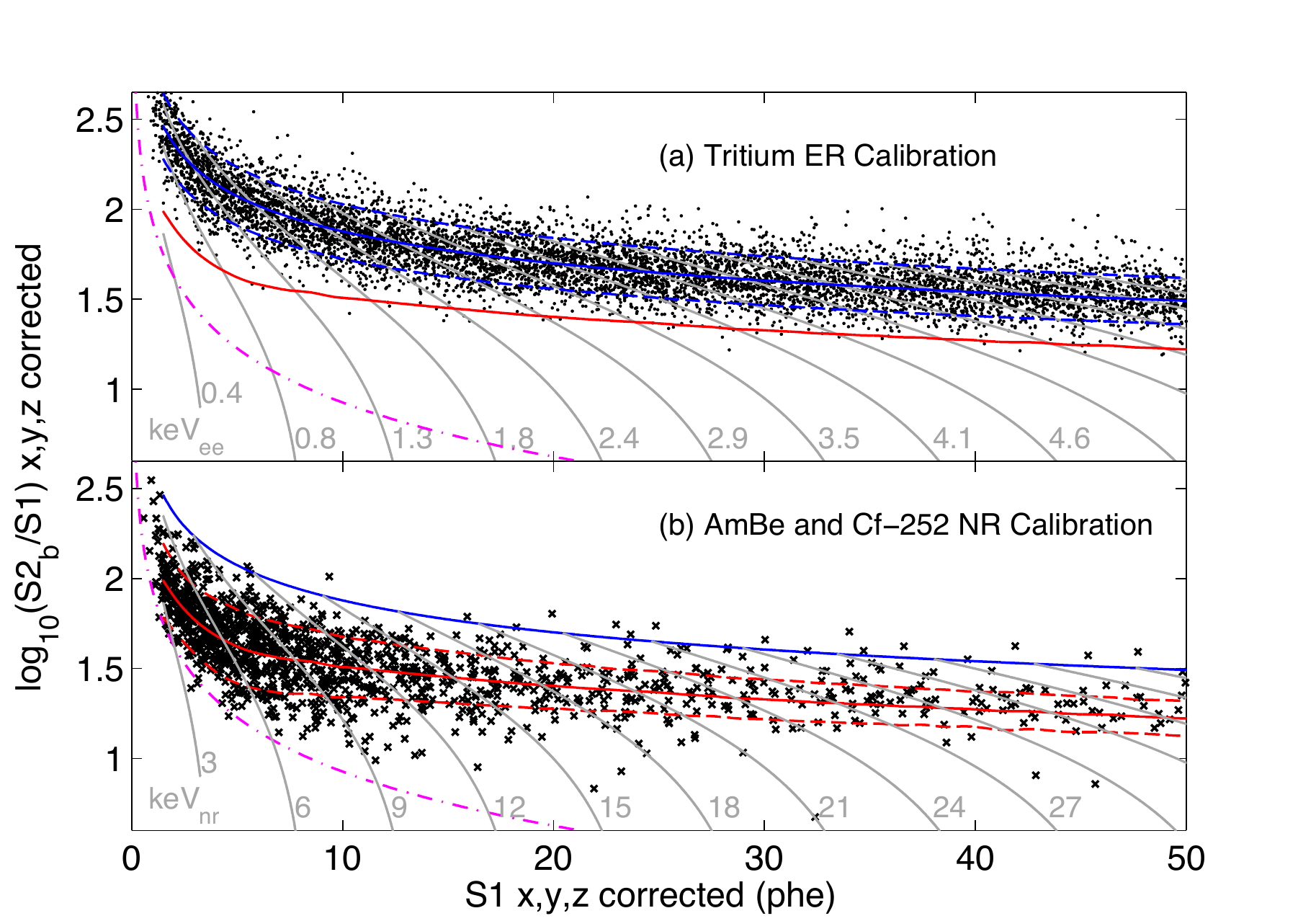}
\caption[]{\label{ERNRbands} Calibrations of detector response in the 118~kg fiducial volume. The ER (tritium, panel $a$) and NR (AmBe and $^{252}$Cf, panel $b$) calibrations are depicted, with the means (solid line) and $\pm 1.28 \sigma$ contours (dashed line). This choice of band width (indicating 10\% band tails) is for presentation only.  Panel $a$ shows fits to the high statistics tritium data, with fits to simulated NR data shown in panel $b$, representing the parameterizations taken forward to the profile likelihood analysis.  The ER plot also shows the NR band mean and vice versa.  Gray contours indicate constant energies using an S1--S2 combined energy scale (same contours on each plot).  The dot-dashed magenta line delineates the approximate location of the minimum S2 cut.}
\end{center}
\end{figure}

Extensive calibrations were acquired with internal ER sources (tritiated methane, $^{83m}$Kr) and NR calibrations were performed with external neutron sources (AmBe, $^{252}$Cf). The ER sources were injected into the xenon gas system and allowed to disperse uniformly, achieving a homogeneous calibration of the active region. In particular, we developed a novel tritiated-methane $\beta^-$ source ($E_{\max} \! \simeq \! 18$~keV) that produces events extending below 1~\keVee, allowing ER band (Fig.~\ref{ERNRbands}) and detection efficiency calibrations (Fig.~\ref{Efficiencies}) with unprecedented accuracy; the tritiated methane is subsequently fully removed by circulating the xenon through the getter.

A $^{83m}$Kr injection was performed weekly to determine the free electron lifetime and the three-dimensional correction functions for photon detection efficiency, which combine the effects of geometric light collection and PMT quantum efficiency (corrected S1 and S2). The 9.4 and 32.1~keV depositions~\cite{Kastens} demonstrated the stability of the S1 and S2 signals in time, the latter confirmed with measurements of the single extracted electron response. $^{131m}$Xe and $^{129m}$Xe (164 and 236~keV de-excitations) afforded another internal calibration, providing a cross-check of the photon detection and electron extraction efficiencies. To model these efficiencies, we employed field- and energy-dependent absolute scintillation and ionization yields from NEST~\cite{Szydagis2013,Szydagis2011,Baudis2013}, which provides an underlying physics model, not extrapolations, where only detector parameters such as photon detection efficiency, electron extraction efficiency and single electron response are inputs to the simulation.  Using a Gaussian fit to the single phe area~\cite{CarlosFahamThesis}, together with the S1 spectrum of tritium events, the mean S1 photon detection efficiency was determined to be $0.14 \pm 0.01$, varying between 0.11 and 0.17 from the top to the bottom of the active region. This is estimated to correspond to 8.8~phe/\keVee{} (electron-equivalent energy) for 122~keV $\gamma$-rays at zero field~\cite{Szydagis2013}. This high photon detection efficiency (unprecedented in a xenon WIMP-search TPC) is responsible for the low threshold and good discrimination observed~\cite{DahlThesis}.

\begin{figure}[h]
\begin{center}
\includegraphics[width=0.48\textwidth,clip]{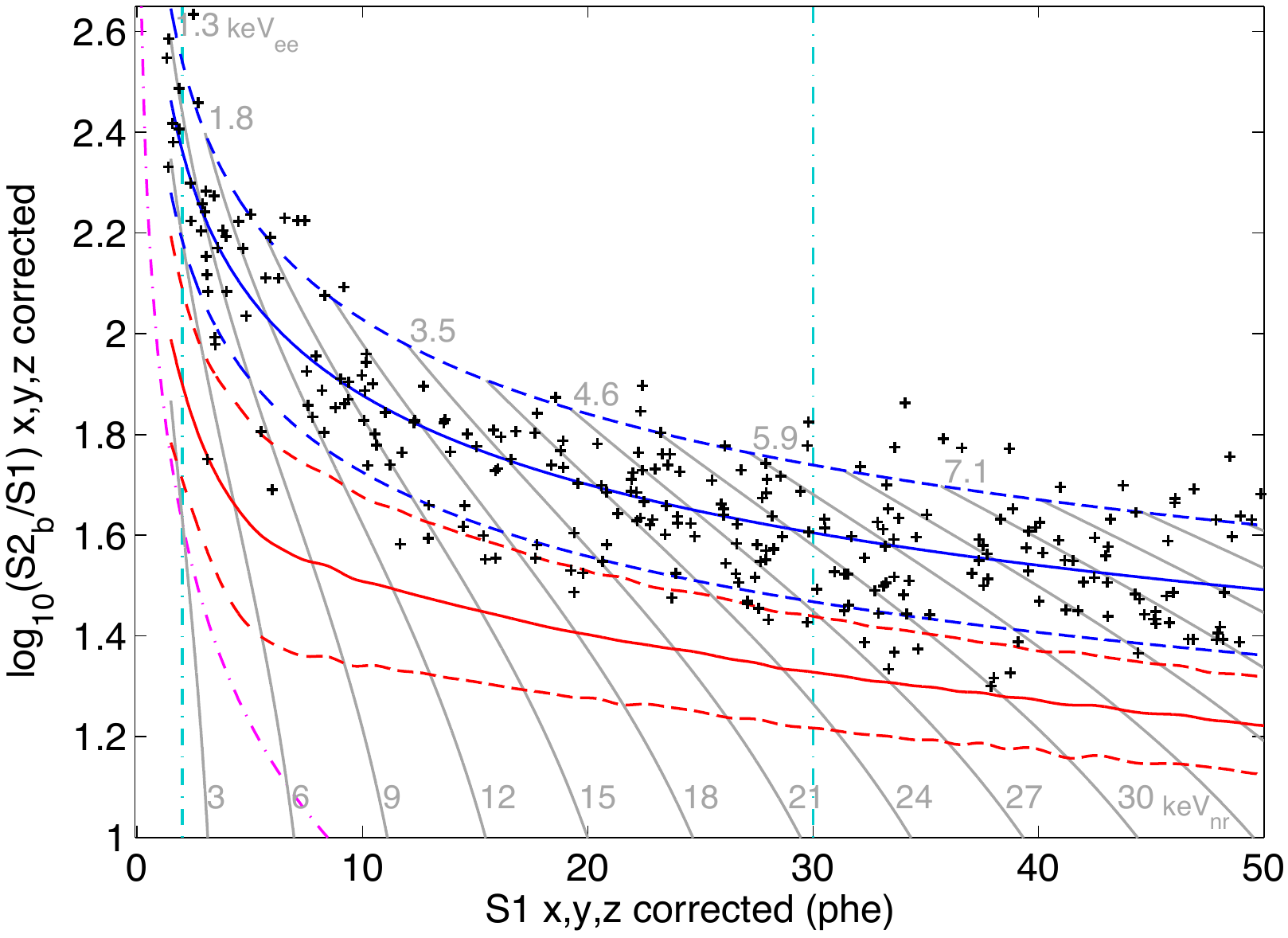}
\caption{\label{WSscatter} The LUX WIMP signal region.  Events in the 118~kg fiducial volume during the 85.3~live-day exposure are shown.  Lines as shown in Fig.~\ref{ERNRbands}, with vertical dashed cyan lines showing the 2-30~phe range used for the signal estimation analysis.}
\end{center}
\end{figure}

Detector response to ER and NR calibration sources is presented in Fig.~\ref{ERNRbands}. Comparison of AmBe data with simulation permits extraction of NR detection efficiency (Fig.~\ref{Efficiencies}), which is in excellent agreement with that obtained using other datasets ($^{252}$Cf and tritium). We describe the populations as a function of S1 (Fig.~\ref{ERNRbands} and Fig.~\ref{WSscatter}), as this provides the dominant component of detector efficiency.  We also show contours of approximated constant-energy~\cite{Sorensen2012}, calculated from a linear combination of S1 and S2~\cite{DahlThesis,SorDahl2011,Szydagis2011} generated by converting the measured pulse areas into original photons and electrons (given their efficiencies).

A parameterization (for S2 at a given S1) of the ER band from the high-statistics tritium calibration is used to characterize the background.  In turn, the NR calibration is more challenging, partly due to the excellent self-shielding of the detector.  Neutron calibrations therefore include systematic effects not applicable to the WIMP signal model, such as multiple-scattering events (including those where scatters occur in regions of differing field) or coincident Compton scatters from AmBe and $^{252}$Cf $\gamma$-rays and (n,$\gamma$) reactions. These effects produce the dispersion observed in data, which is well modeled in our simulations (in both band mean and width, verifying the simulated energy resolution), and larger than that expected from WIMP scattering. Consequently, these data cannot be used directly to model a signal distribution. For different WIMP masses, simulated S1 and S2 distributions are obtained, accounting for their unique energy spectra.

The ratio of \keVee{} to nuclear recoil energy (\keVnr) relies on both S1 and S2, using the conservative technique presented in~\cite{SorDahl2011} (Lindhard with $k=0.110$, compared to the default Lindhard value of 0.166 and the implied best-fit value of 0.135 from~\cite{SorDahl2011}). NR data are consistent with an energy-dependent, non-monotonic reduced light yield with respect to zero field~\cite{Plante2011} with a minimum of~0.77 and a maximum of 0.82 in the range 3--25~\keVnr{}~\cite{Szydagis2013} (compared with 0.90-0.95 used by previous xenon experiments for significantly higher electric fields~\cite{ZIII,Xe100}). This is understood to stem from additional, anti-correlated portioning into the ionization channel.

\begin{table}[h]
\caption{Predicted background rates in the fiducial volume (0.9--5.3~\keVee)~\cite{LUXBGPaper2013Malling}. We show contributions from the $\gamma$-rays of detector components (including those cosmogenically activated), the time-weighted contribution of activated xenon, $^{222}$Rn (best estimate 0.2~mDRUee from $^{222}$Rn chain measurements) and $^{85}$Kr.  The errors shown are both from simulation statistics and those derived from the rate measurements of time-dependent backgrounds. 1~mDRU$_{\mathrm{ee}}$ is $10^{-3}$ events/keV$_{\mathrm{ee}}$/kg/day.}
\label{BGtable}
\begin{tabular}{| c | c |} 
\hline  
Source & Background rate, mDRU$_{\mathrm{ee}}$ \\
\hline 
$\gamma$-rays & $1.8 \pm 0.2_\textrm{{stat}} \pm 0.3_\textrm{{sys}}$ \\ 
$^{127}$Xe & $0.5 \pm 0.02_\textrm{{stat}} \pm 0.1_\textrm{{sys}}$ \\ 
$^{214}$Pb & 0.11--0.22 (90\% C.\,L.) \\ 
$^{85}$Kr & $0.13 \pm 0.07_\textrm{{sys}}$ \\ 
\hline 
\hline 
Total predicted & $2.6 \pm 0.2_\textrm{{stat}} \pm 0.4_\textrm{{sys}}$\\
Total observed & $3.6 \pm 0.3_\textrm{{stat}}$ \\ 
\hline
\end{tabular} 
\end{table}

The observed ER background in the range 0.9--5.3~\keVee{} within the fiducial volume was $3.6 \pm 0.3$~mDRU$_{\mathrm{ee}}$ averaged over the WIMP search dataset (summarized in Table~\ref{BGtable}).  Backgrounds from detector components were controlled through a material screening program at the Soudan Low-Background Counting Facility (SOLO) and the LBNL low-background counting facility~\cite{LUXPMTPaper, CarlosFahamThesis, TiPaper}.  Krypton as a mass fraction of xenon was reduced from 130~ppb in the purchased xenon to 4~ppt using gas charcoal chromatography~\cite{krRemoval}.  

Radiogenic backgrounds were extensively modeled using LUXSim, with approximately 73\% of the low-energy $\gamma$-ray background originating from the materials in the R8778~PMTs and the rest from other construction materials.  This demonstrated consistency between the observed $\gamma$-ray energy spectra and position distribution~\cite{LUXBGPaper2013Malling}, and the expectations based on the screening results and the independent assay of the natural Kr concentration of $3.5 \pm 1$~ppt (g/g) in the xenon gas~\cite{dobi} where we assume an isotopic abundance of $^{85}$Kr/$^{nat}$Kr $\sim2\times10^{-11}$~\cite{ChangAPS,LUXBGPaper2013Malling}.  Isotopes created through cosmogenic production were also considered, including measured levels of $^{60}$Co in Cu components.  {\it In situ} measurements determined additional intrinsic background levels in xenon from $^{214}$Pb (from the $^{222}$Rn decay chain) \cite{KG_TAUP}, and cosmogenically-produced $^{127}$Xe ($T_{1/2} = 36.4$~days), $^{129m}$Xe ($T_{1/2} = 8.9$~days), and $^{131m}$Xe ($T_{1/2} = 11.9$~days).  The rate from $^{127}$Xe in the WIMP search energy window is estimated to decay from 0.87~mDRU$_{\mathrm{ee}}$ at the start of the WIMP search dataset to 0.28~mDRU$_{\mathrm{ee}}$ at the end, with late-time background measurements being consistent with those originating primarily from the long-lived radioisotopes.

The neutron background in LUX is predicted from detailed detector BG simulations to produce 0.06 single scatters with S1 between 2 and 30~phe in the 85.3~live-day dataset. This was considered too low to  include in the PLR. The value was constrained by multiple-scatter analysis in the data, with a conservative 90\% upper C.L.\ placed on the number of expected neutron single scatters of 0.37 events.

We observed 160~events between 2~and 30~phe (S1) within the fiducial volume in 85.3~live-days of search data (shown in Fig.~\ref{WSscatter}), with all observed events being consistent with the predicted background of electron recoils.  The average discrimination (with 50\% NR acceptance) for S1 from 2-30~phe is $99.6 \pm 0.1 \%$, hence $0.64 \pm 0.16$~events from ER leakage are expected below the NR mean, for the search dataset.  The spatial distribution of the events matches that expected from the ER backgrounds in full detector simulations.  We select the upper bound of 30~phe (S1) for the signal estimation analysis to avoid additional background from the 5~\keVee{} x-ray from $^{127}$Xe.

Confidence intervals on the spin-independent WIMP-nucleon cross section are set using a profile likelihood ratio (PLR) test statistic~ \cite{Cowan}, exploiting the separation of signal and background distributions in four physical quantities: radius, depth, light (S1), and charge (S2). The fit is made over the parameter of interest plus three Gaussian-constrained nuisance parameters which encode uncertainty in the rates of $^{127}$Xe, $\gamma$-rays from internal components and the combination of $^{214}$Pb and $^{85}$Kr.  The distributions, in the observed quantities, of the four model components are as described above and do not vary in the fit: with the non-uniform spatial distributions of $\gamma$-ray backgrounds and x-ray lines from $^{127}$Xe obtained from energy-deposition simulations~\cite{LUXBGPaper2013Malling}. The PLR operates within the fiducial region but the spatial background models were validated using data from outside the fiducial volume.

The energy spectrum of WIMP-nucleus recoils is modeled using a standard isothermal Maxwellian velocity distribution~\cite{Savage}, with $v_{0} = 220\;\mathrm{km/s}$; $v_{\mathrm{esc}} = 544\;\mathrm{km/s}$; $\rho_{0} = 0.3\;\mathrm{GeV}/\mathrm{cm}^{3}$; average Earth velocity of $245\;\mathrm{km\;s^{-1}}$, and Helm form factor \cite{Helm,LandS}. We conservatively model no signal below 3.0~\keVnr{} (the lowest energy for which a direct light yield measurement exists~\cite{Manzur2010,Plante2011}, whereas indirect evidence of charge yield exists down to 1~\keVnr{}~\cite{Sorensen2010}).  We do not profile the uncertainties in NR yield, assuming a model which provides excellent agreement with LUX data (Fig.~\ref{Efficiencies} and Fig.~\ref{Qy}), in addition to being conservative compared to past works~\cite{Szydagis2013}.  We also do not account for uncertainties in astrophysical parameters, which are beyond the scope of this work (but are discussed in~\cite{McCabe}).  Signal models in S1 and S2 are obtained for each WIMP mass from full simulations.  

\begin{figure}[h]
\begin{center}
\includegraphics[width=0.48\textwidth,clip]{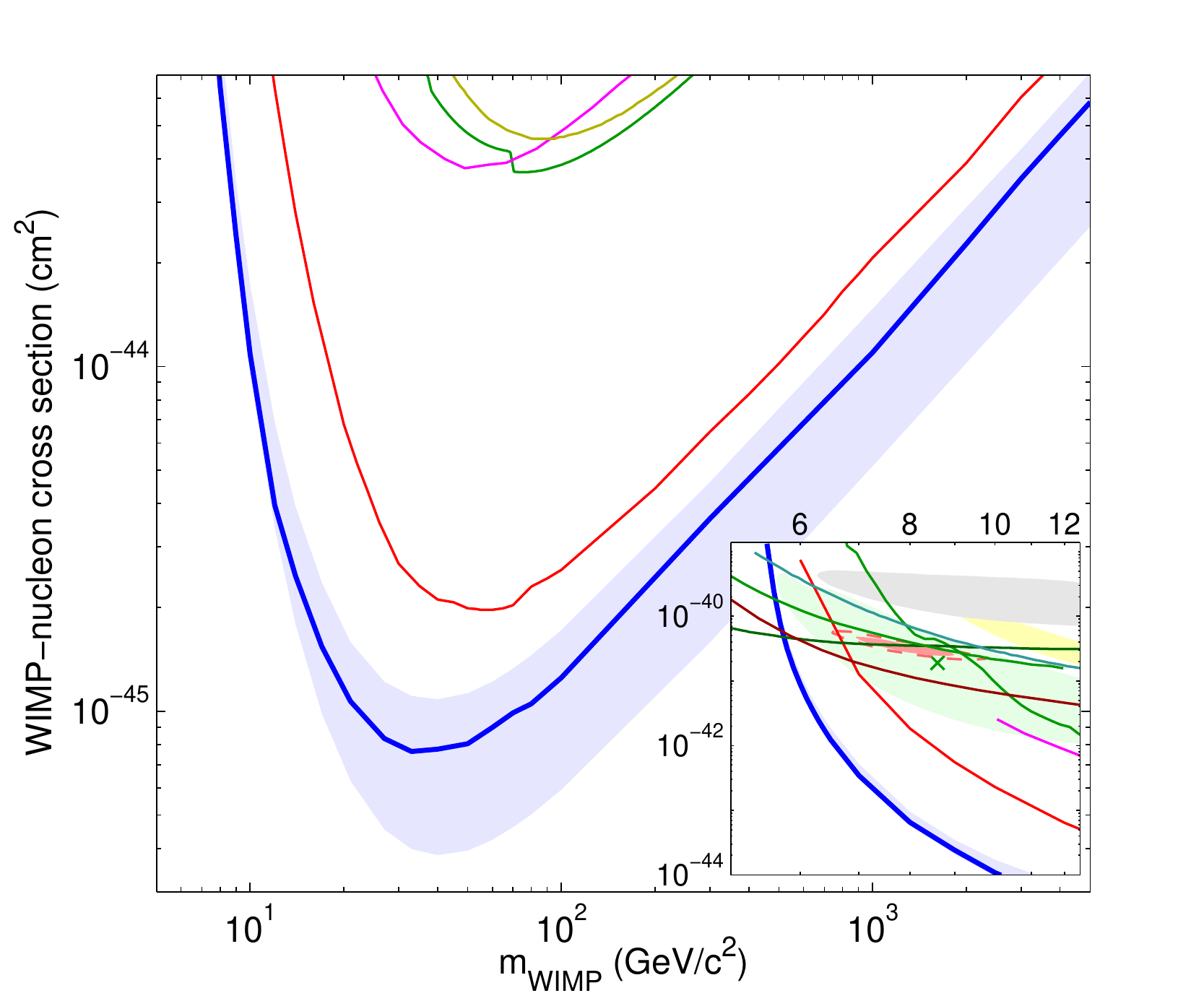}
\caption{\label{limitPlot} The LUX 90\% confidence limit on the spin-independent elastic WIMP-nucleon cross section (blue), together with the ${\pm}1\sigma$ variation from repeated trials, where trials fluctuating below the expected number of events for zero BG are forced to 2.3 (blue shaded). We also show Edelweiss~II~\cite{edelweiss} (dark yellow line), CDMS~II~\cite{CDMSLatestResults} (green line), ZEPLIN-III~\cite{ZIII} (magenta line), CDMSlite~\cite{CDMSlite} (dark green line), XENON10 S2-only~\cite{PeterXe10S2only} (brown line), SIMPLE~\cite{SIMPLE} (light blue line) and XENON100 100~live-day~\cite{Xe1002011} (orange line), and 225~live-day~\cite{Xe100} (red line) results. The inset (same axis units) also shows the regions measured from annual modulation in CoGeNT~\cite{CoGENTLatestResults} (light red, shaded), along with exclusion limits from low threshold re-analysis of CDMS~II data~\cite{CDMS_2keV} (upper green line), 95\% allowed region from CDMS~II silicon detectors~\cite{CDMS_SiResults} (green shaded) and centroid (green x), 90\% allowed region from CRESST II~\cite{CRESST} (yellow shaded) and DAMA/LIBRA allowed region~\cite{DAMA} interpreted by~\cite{DAMA_Savage} (grey shaded).  Results sourced from DMTools~\cite{dmtools}.}
\end{center}
\end{figure}

The observed PLR for zero signal is entirely consistent with its simulated distribution, giving a p-value for the background-only hypothesis of 0.35. The 90\% C.\,L.\, upper limit on the number of expected signal events ranges, over WIMP masses, from 2.4 to~5.3.  A variation of one standard deviation in detection efficiency shifts the limit by an average of only 5\%.  The systematic uncertainty in the position of the NR band was estimated by averaging the difference between the centroids of simulated and observed AmBe data in log(S2$_b$/S1).  This yielded an uncertainty of 0.044 in the centroid, which propagates to a maximum uncertainty of 25\% in the high mass limit.

The 90\% upper C.\,L.\, cross sections for spin-independent WIMP models are thus shown in Fig.~\ref{limitPlot} with a minimum cross section of $7.6 {\times} 10^{-46}$~cm$^2$ for a WIMP mass of 33~GeV/c$^2$. This represents a significant improvement over the sensitivities of earlier searches~\cite{Xe100, CDMSLatestResults, CoGENTLatestResults, ZIII}.  The low energy threshold of LUX permits direct testing of low mass WIMP hypotheses where there are potential hints of signal~\cite{CDMSLatestResults,CoGENTLatestResults,CRESST,DAMA}. These results do not support such hypotheses based on spin-independent isospin-invariant WIMP-nucleon couplings and conventional astrophysical assumptions for the WIMP halo, even when using a conservative interpretation of the existing low-energy nuclear recoil calibration data for xenon detectors.

LUX will continue operations at SURF during 2014 and 2015.  Further engineering and calibration studies will establish the optimal parameters for detector operations, with potential improvements in applied electric fields, increased calibration statistics, decaying backgrounds and an instrumented water tank veto further enhancing the sensitivity of the experiment.  Subsequently, we will complete the ultimate goal of conducting a blinded 300~live-day WIMP search further improving sensitivity to explore significant new regions of WIMP parameter space. 

This work was partially supported by the U.S. Department of Energy (DOE) under award numbers DE-FG02-08ER41549, DE-FG02-91ER40688, DE-FG02-95ER40917, DE-FG02-91ER40674, DE-NA0000979, DE-FG02-11ER41738, DE-SC0006605, DE-AC02-05CH11231, DE-AC52-07NA27344, and DE-FG01-91ER40618; the U.S. National Science Foundation under award numbers PHYS-0750671, PHY-0801536, PHY-1004661, PHY-1102470, PHY-1003660, PHY-1312561, PHY-1347449; the Research Corporation grant RA0350; the Center for Ultra-low Background Experiments in the Dakotas (CUBED); and the South Dakota School of Mines and Technology (SDSMT). LIP-Coimbra acknowledges funding from Funda\c{c}\~{a}o para a Ci\^{e}ncia e Tecnologia (FCT) through the project-grant CERN/FP/123610/2011. Imperial College and Brown University thank the UK Royal Society for travel funds under the International Exchange Scheme (IE120804). The UK groups acknowledge institutional support from Imperial College London, University College London and Edinburgh University, and from the Science \& Technology Facilities Council for Ph.\,D.\, studentship ST/K502042/1 (AB). The University of Edinburgh is a charitable body, registered in Scotland, with registration number SC005336.  This research was conducted using computational resources and services at the Center for Computation and Visualization, Brown University.

We acknowledge the work of the following engineers who played important roles during the design, construction, commissioning, and operation phases of LUX: S. Dardin from Berkeley, B. Holbrook, R. Gerhard, and J. Thomson from University of California, Davis; and G. Mok, J. Bauer, and D. Carr from Lawrence Livermore National Laboratory.
We gratefully acknowledge the logistical and technical support and access to laboratory infrastructure provided to us by the Sanford Underground Research Facility (SURF) and its personnel at Lead, South Dakota. SURF was developed by the South Dakota Science and Technology authority, with an important philanthropic donation from T. Denny Sanford, and is operated by Lawrence Berkeley National Laboratory for the Department of Energy, Office of High Energy Physics.

\clearpage
\begin{widetext}

\appendix
\section{Supplementary Material}
This document contains supplementary material in support the main article.  We show details of:
\begin{itemize}
\item Figure~\ref{Qy} -- the matching of AmBe MC simulations and data in the ionization channel.
\item Table~\ref{cutcounttable} -- the number of events in the WIMP search dataset, following the application of each set of cuts.
\item Figures~\ref{highEbg} and \ref{lowEbg} -- the observed and expected background energy spectra at high and low energy.
\item Figure~\ref{discrimination} -- the discrimination/leakage fraction of ER to NR signals as a function of S1.
\item Figure~\ref{neutronSimsComparison} - agreement between mean and width for neutron calibration simulations and data.
\item Figure~\ref{tritium_pde_eee} - tritium S1 spectrum (showing best-fit photon detection efficiency) and activated xenon peaks in S2 versus S1 space (showing best-fit for photon detection efficiency, electron extraction efficiency and single electron response).
\end{itemize}

\begin{figure}[h]
\begin{center}
\includegraphics[width=0.7\textwidth,clip]{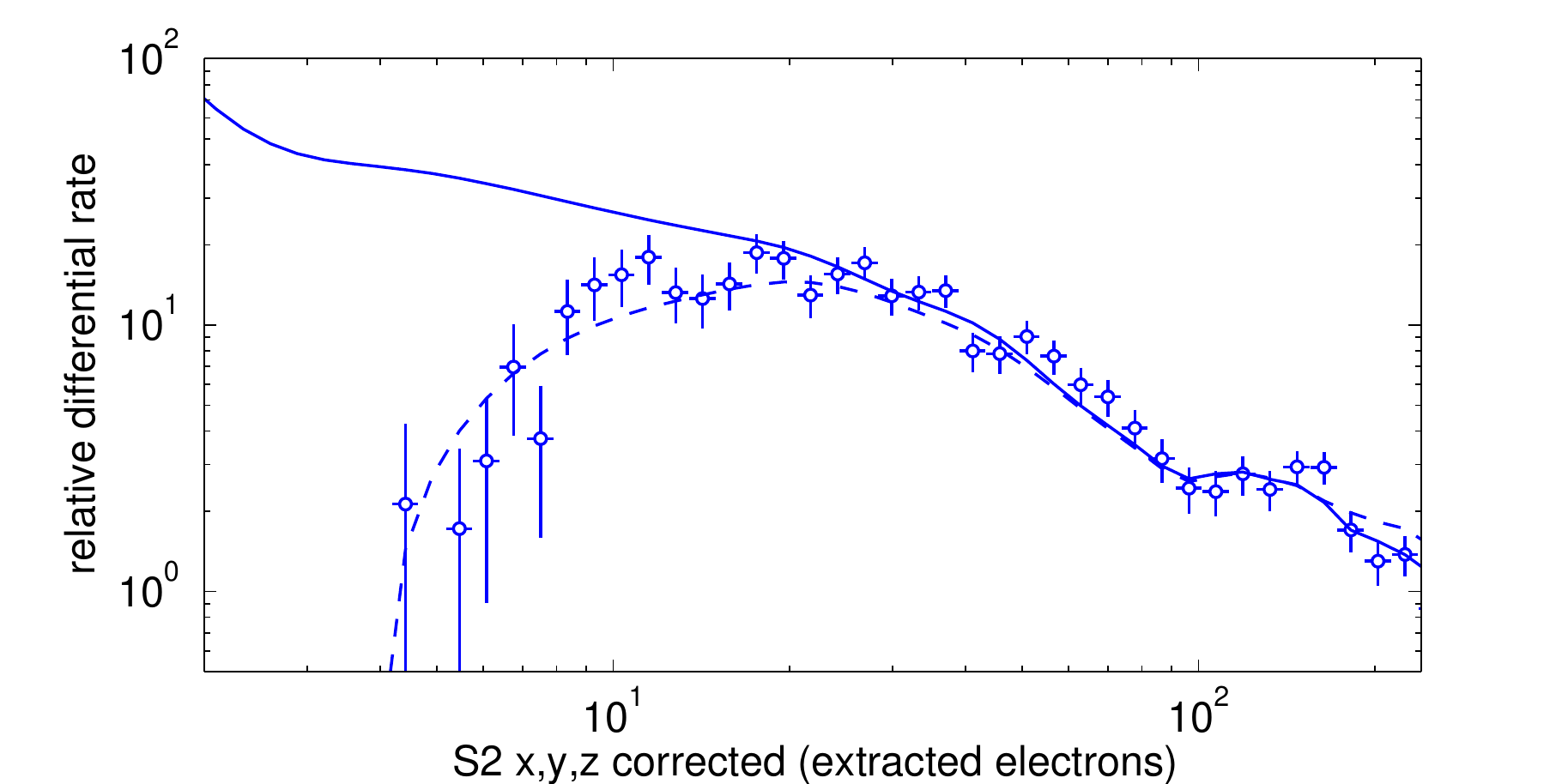}
\caption{\label{Qy}  Comparison of single scatter Am-Be calibration nuclear recoil events (blue data points) with Monte Carlo simulations. The blue solid curve shows the full simulated spectrum, whereas the blue dashed curve shows the expected spectrum when the detector efficiency (Fig~1 main article) is applied. Agreement is found to below the applied S2 pulse threshold (200~phe), demonstrating that not only the light yield but also the charge yield data is well described by the NEST simulation, used in the PLR to model signal as a function of WIMP mass.}
\end{center}
\end{figure}

\begin{figure}[h]
\begin{center}
\includegraphics[width=0.48\textwidth,clip]{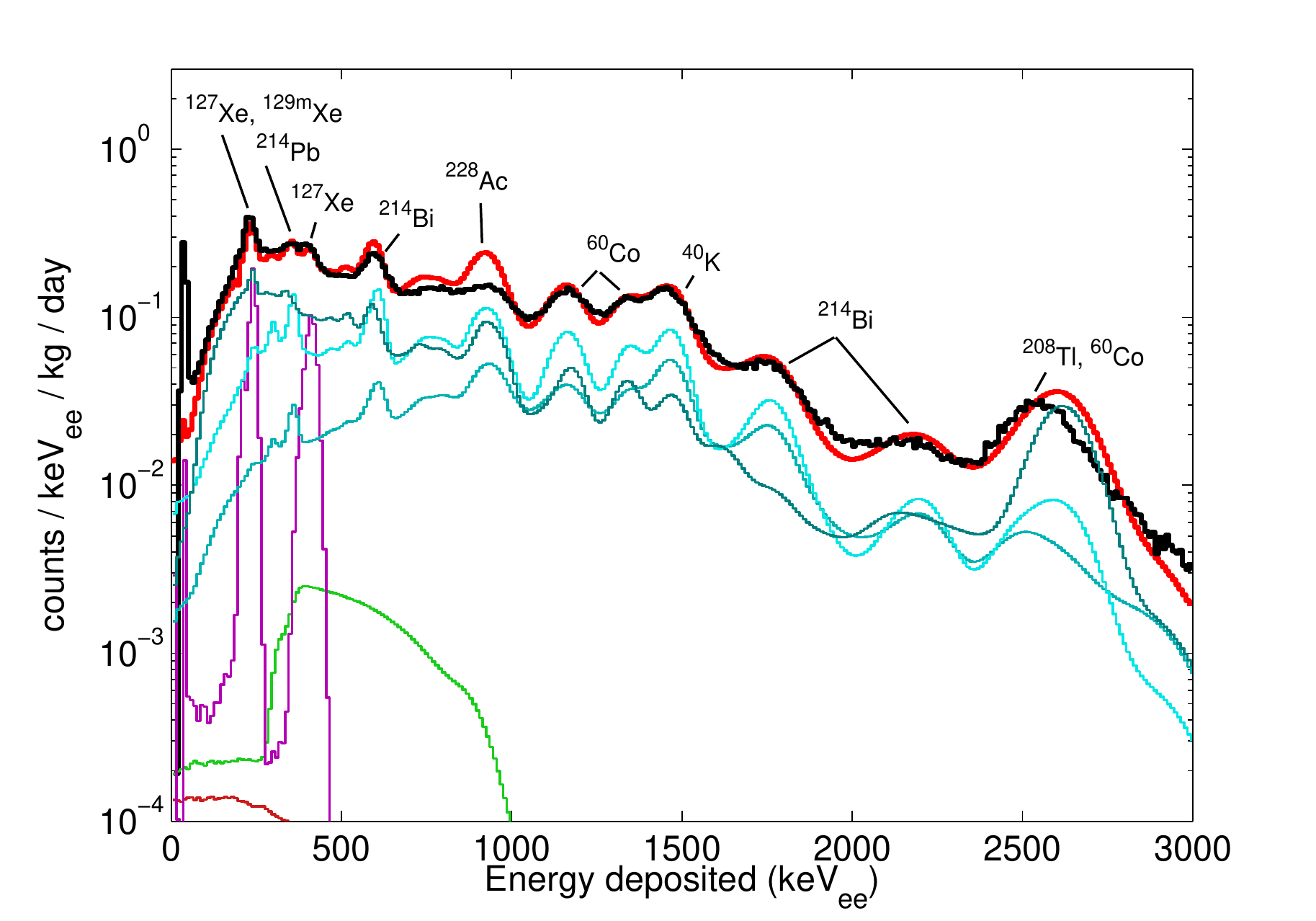}
\includegraphics[width=0.48\textwidth,clip]{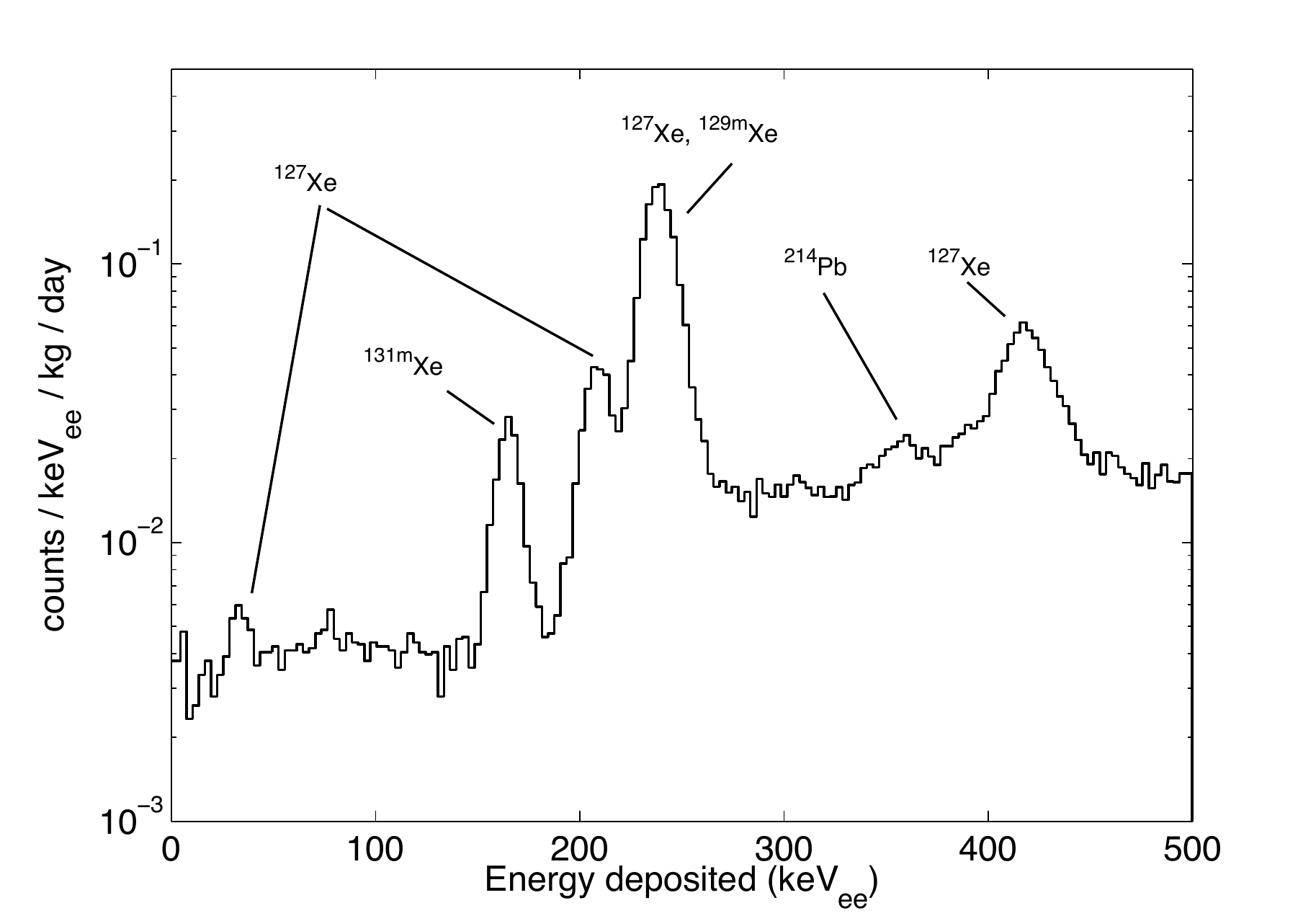}
\caption{\label{highEbg}  Left:  Measured gamma energy spectrum in the 225~kg (2~cm cut top and bottom to remove events grid wires or irregular field regions) LUX drift region (black). Measured spectrum includes both single and multiple scatter events, and is reconstructed from combined S1 and S2 signals. No fiducial cuts are used. The high-energy spectrum from simulation (red) is also shown based on best-fit parameters with measured data. Simulations feature gammas generated from $^{238}$U, $^{232}$Th, $^{40}$K, and $^{60}$Co decays, spread over the top (blue), bottom (dark blue), and side (light blue) construction materials adjoining the active region, as well as activated xenon (purple), $^{85}Kr$ (red), and $^{214}Pb$ (green) evenly distributed in the bulk. The best-fit spectrum 
was matched to data over 13~slices in depth, for energies $>$500~keVee.~\cite{LUXBGPaper2013Malling}\\
Right:  Measured background distribution for the fiducial region in the range 0--500~keV$_{ee}$ with the main peaks identified.}
\end{center}
\end{figure}

\begin{figure}[h]
\begin{center}
\includegraphics[width=0.7\textwidth,clip]{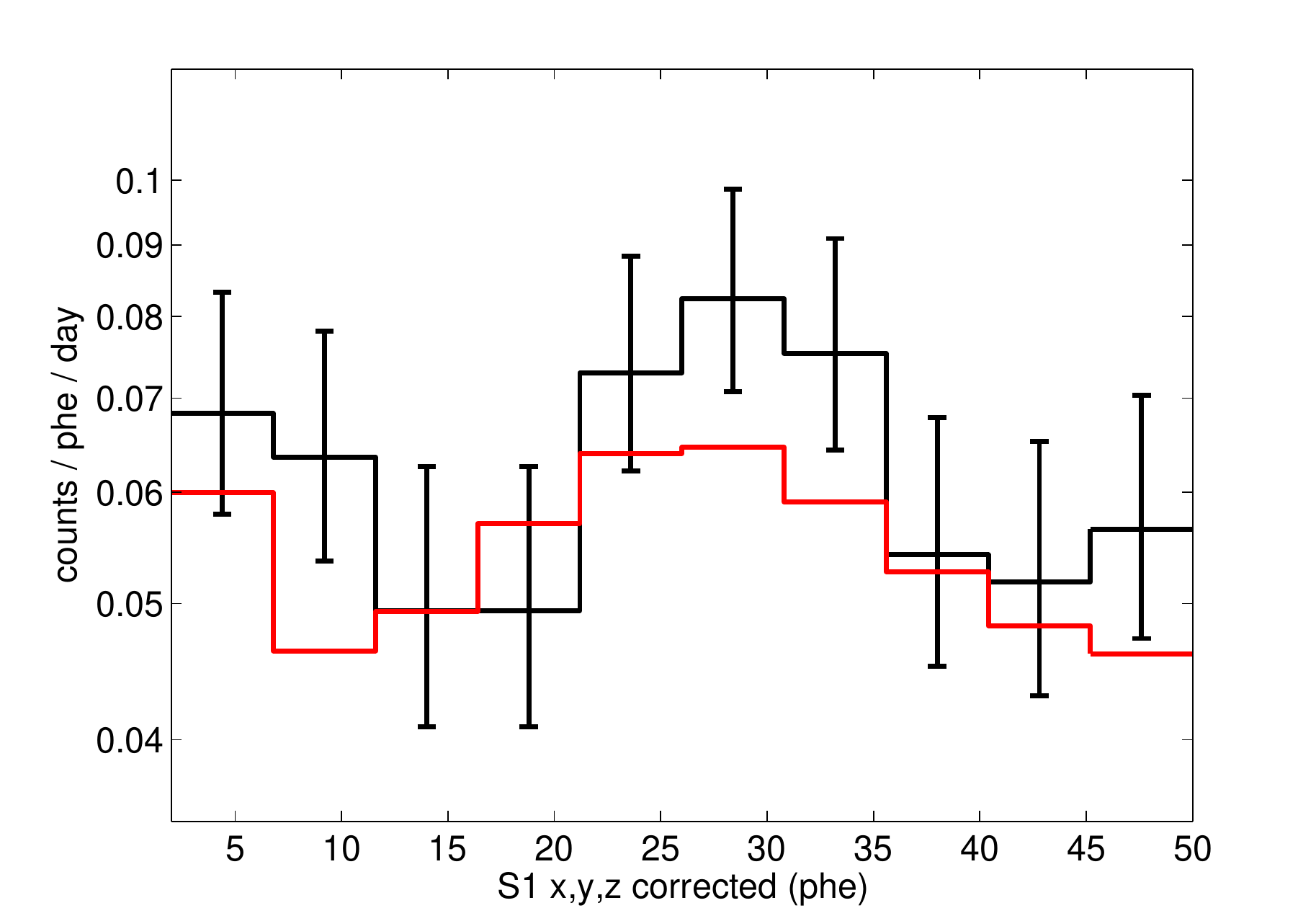}
\caption{\label{lowEbg}  S1 distribution of low-energy backgrounds within the 118~kg WIMP search fiducial volume, ranging from 2--50~phe (black). Data shown includes all WIMP search selection cuts. Simulation predictions based on models of gamma, $^{127}$Xe, $^{214}$Pb, and $^{85}$Kr low-energy background contributions are shown for comparison (red). Gamma, $^{127}$Xe and $^{214}$Pb low-energy model predictions are extrapolated from independent high-energy measurements. $^{85}$Kr rates are estimated based on the direct measurement of the Kr content of LUX xenon, with an assumed $2\times10^{-11}$~$^{85}$Kr~/~nat~Kr. Count expectations are based on an 85~live-day exposure.~\cite{LUXBGPaper2013Malling}}
\end{center}
\end{figure}

\begin{figure}[h]
\begin{center}
\includegraphics[width=0.7\textwidth,clip]{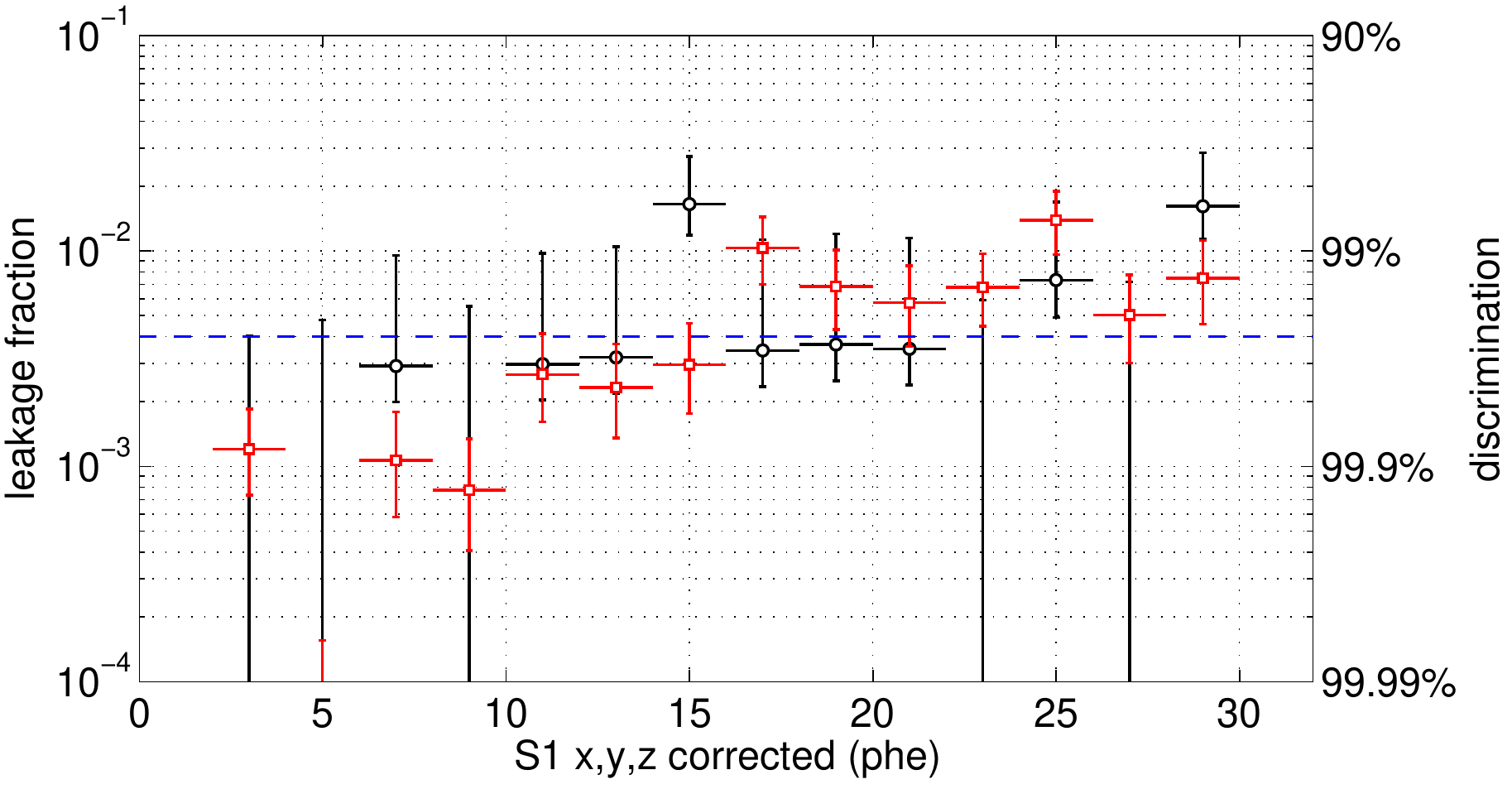}
\caption{\label{discrimination}  Plot showing the leakage fraction (discrimination level) between electron and nuclear recoil populations, with 50\% nuclear recoil acceptance (as calculated from flat-in-energy NR simulations), measured with the high-statistics tritium data.  We show the leakage from counting events in the dataset (black circles) and from projections of Gaussian fits to the electron recoil population (red squares).  An upper limit is shown for S1 bins without events.  The blue dashed line indicates the total leakage fraction, 0.004, in the S1 range 2-30~phe.  The leakage fraction is not used directly in the estimation of the WIMP signal.}
\end{center}
\end{figure}

\begin{table}[h]
\begin{center}
\begin{tabular}{| c | c |} 
\hline  
Cut & Events Remaining\\ 
\hline 
\hline 
all triggers & 83,673,413 \\ 
\hline 
detector stability & 82,918,902 \\ 
\hline 
single scatter & 6,585,686 \\ 
\hline 
S1 energy ($2-30~$phe) & 26,824 \\
\hline
S2 energy ($200-3300~$phe) & 20,989 \\
\hline
single electron background & 19,796 \\
\hline
fiducial volume & 160 \\
\hline

\end{tabular} 
\caption{\label{cutcounttable}Number of events remaining after each analysis cut. All of these cuts are commutative, the order indicating the order in which the cuts are applied in the analysis. Detector stability cuts remove periods of live-time when the liquid level, gas pressure, or grid voltages were out of nominal ranges. The single scatter cut keeps only events containing one S1 and one S2 pulse, representative of expected elastic scattering of WIMPs. This cut removes multiple scatter events, S1-only, S2-only events, two event windows that overlap in time, and trigger windows that contain no S1 or S2 signals. S1 and S2 energy cuts keep only those events in the WIMP search energy range. Additionally, the S2 energy threshold of 200~phe removes single-extracted-electron-type events and events with unreliable position reconstruction. Periods of live-time with high rates of single electron backgrounds are then removed. The fiducial volume cut selects only those events with reconstructed radius less than 18~cm, and electron drift time between 38 and 305~$\mu$s. The final number of events in the WIMP search profile likelihood is 160.  A more detailed description of the cuts is provided in~\cite{LUXRun03}.} 
\end{center}
\end{table}

\begin{figure}[h]
\begin{center}
\includegraphics[width=0.48\textwidth,clip]{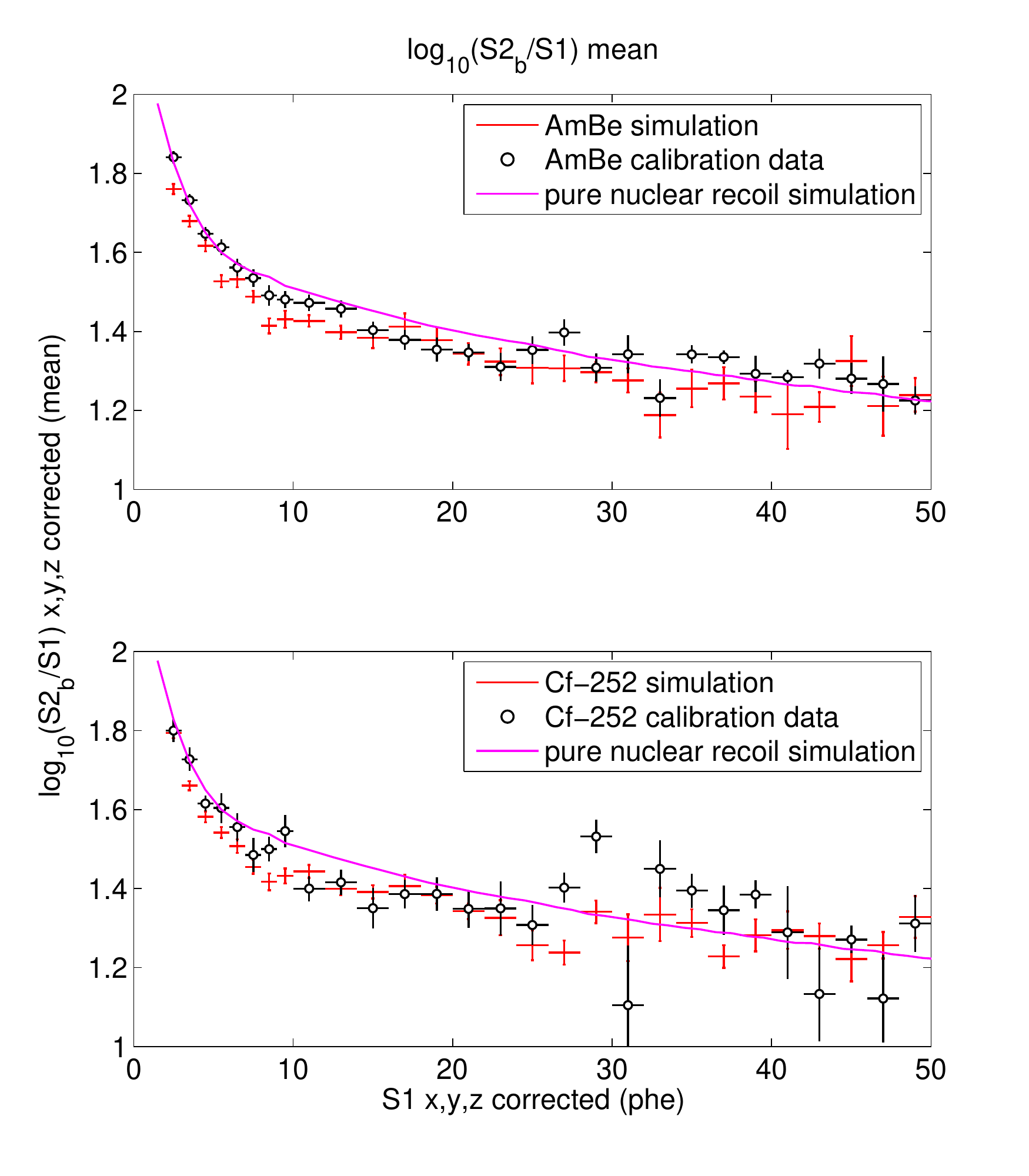}
\includegraphics[width=0.48\textwidth,clip]{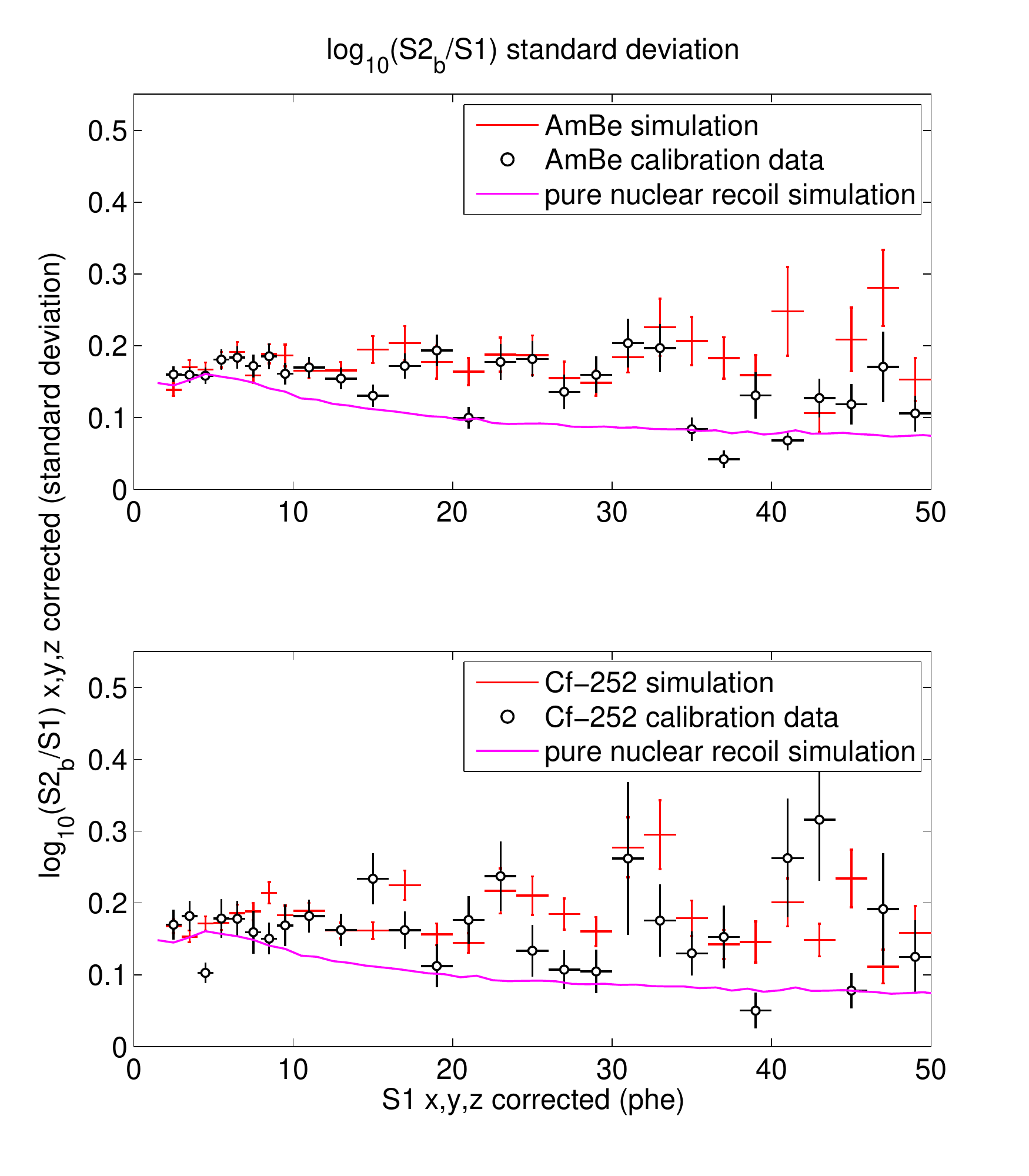}
\caption{\label{neutronSimsComparison} Plots showing the mean (left) and width (right) of the NR band measured from the AmBe (top) and $^{252}$Cf (bottom) calibrations compared with that from neutron calibrations simulations and that from pure NR simulations.  This demonstrates the good agreement between simulation and data with the replication of the neutron plus gamma and neutron-X components to the calibration in addition to the WIMP-like pure NR component.}
\end{center}
\end{figure}

\begin{figure}[h]
\begin{center}
\includegraphics[width=0.48\textwidth,clip]{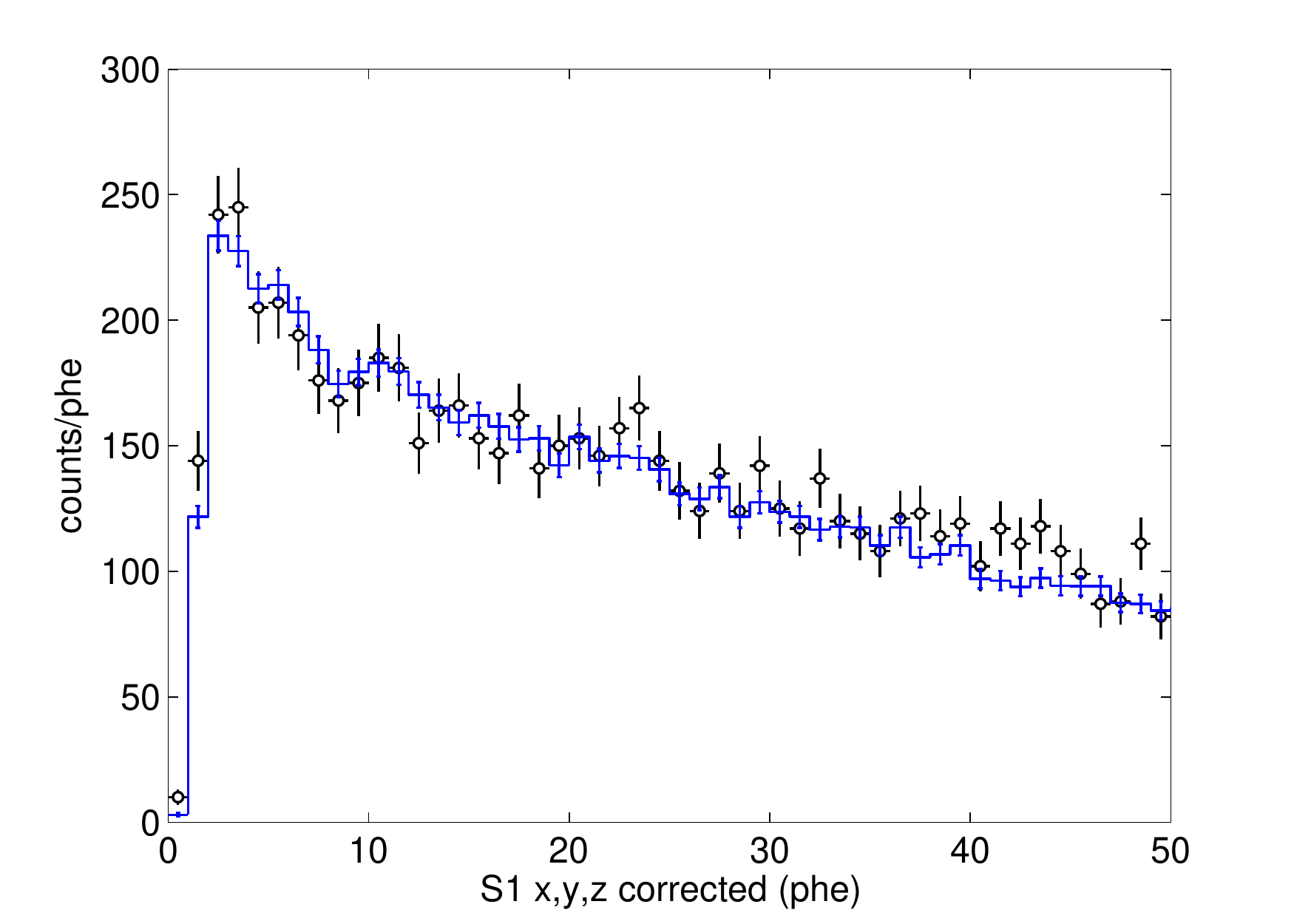}
\includegraphics[width=0.48\textwidth,clip]{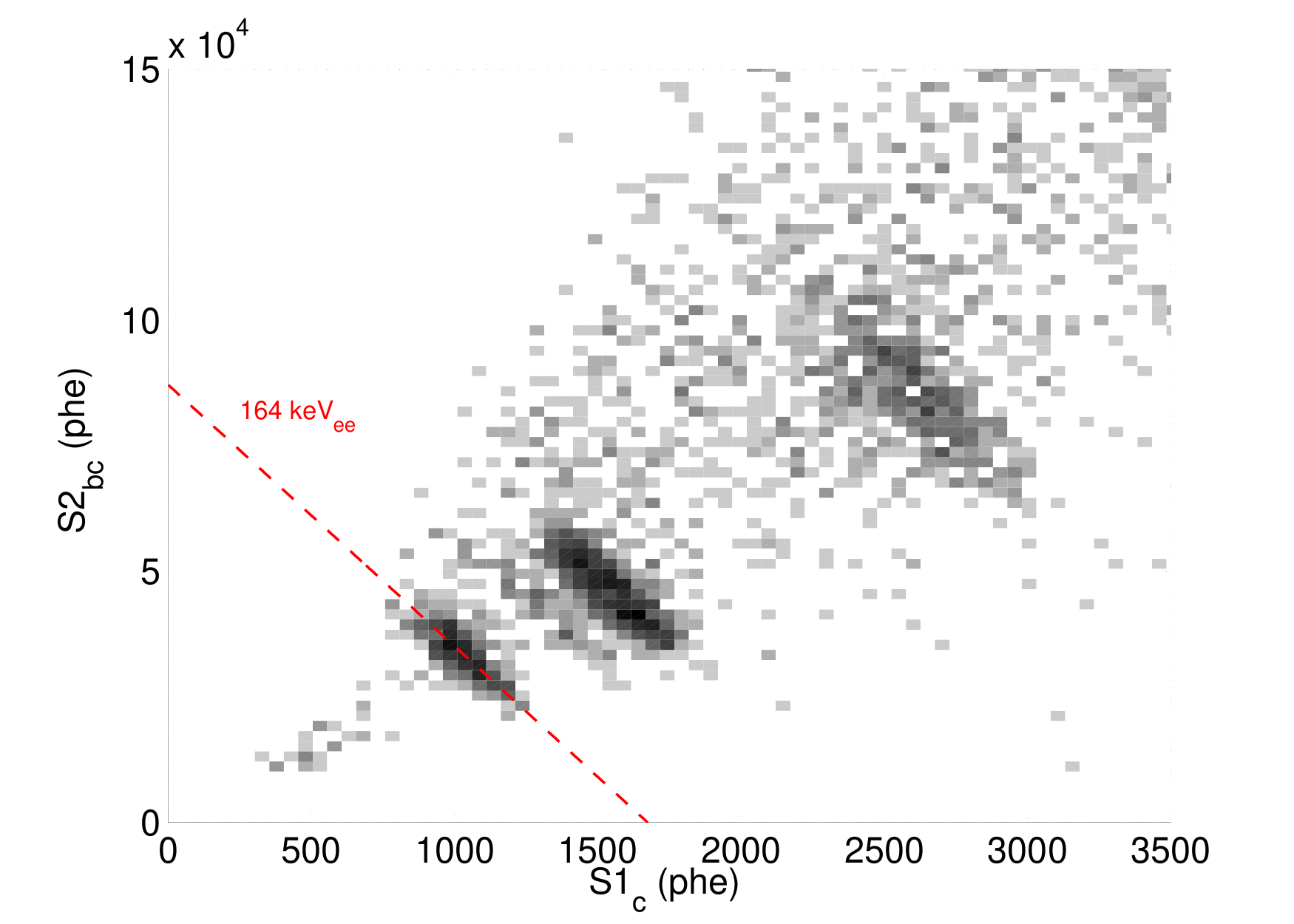}
\caption{\label{tritium_pde_eee} Tritium S1 spectrum (left), comparing data (black) with simulations (blue) for the best-fit photon detection efficiency of $0.14 \pm 0.01$.  Plot showing the activated xenon peaks in S2 versus S1 space (right), where the red dashed line shows the combined energy transformation using the best-fit photon detection efficiency, electron extraction efficiency and measured single electron response.} 
\end{center}
\end{figure}

\end{widetext}

\end{document}